\documentstyle[twocolumn,prl,aps]{revtex}
\input psfig
\begin{document}
\draft
\title{Exact solution for two interacting electrons on artificial atoms 
and molecules in solids}
\author{Amnon Aharony$^1$, Ora Entin-Wohlman$^1$ and Yoseph Imry$^2$}
\address{$^1$ School of Physics and Astronomy, 
Sackler Faculty of Exact Sciences,
Tel Aviv University, Tel Aviv 69978, Israel}
\address{$^2$ Department of Condensed Matter Physics, Weizmann Institute of 
Science,
Rehovot 76100, Israel}
\date{\today}
\maketitle
\begin{abstract}
We present a general scheme for finding the exact eigenstates of
two electrons, with on--site repulsive potentials $\{ U_i \}$, on
$I$ impurities in a 
macroscopic crystal.
The model describes impurities in doped semiconductors
and artificial ``molecules" in quantum dots.
For quantum dots, 
the energy cost for adding two electrons is bounded
by the single--electron spectrum, and does not diverge when
$U_i \rightarrow \infty$,
implying limitations on the validity of the 
Coulomb blockade picture.
Analytic applications 
on a one--dimensional chain yield
quantum delocalization
and magnetic transitions.
\end{abstract}

There has been much recent interest in the effects of interactions
on the localization of electrons in disordered systems \cite{belitz}.
An important question, triggered by recent experiments in two dimensions
\cite{kravchenko}, concerns the possibility that the interactions help
to delocalize the electrons, yielding a metal--insulator transition.
For two interacting electrons in a random potential, this
possibility was supported by Shepelyansky's numerical work
\cite{shepelyansky}, by a Thouless type block--scaling argument \cite{imry}
and by further numerical scaling results 
\cite{vonoppen,ortuno}.
The latter took advantage of the fact that, for two electrons
on an $N$--site lattice with on--site repulsion,
it is sufficient
to study the two--particle Green's function only in the $N$--dimensional
Hilbert space of the doubly occupied sites, instead of the full 
$N^2$--dimensional
space \cite{vonoppen}.
In the present paper we generalize such considerations for the dilute
case, in which only $I$ out of the $N$ sites are replaced by
impurities. These impurities may have different single--electron
(1e) energies $\{\epsilon_i\}$ and on--site
electron--electron (e--e) interactions $\{U_i\}$.
As we show, the exact eigenenergies of the two electrons are found
from the eigenvalues of a (small) $I \times I$ matrix, which involves the
eigenfunctions and the eigenvalues of the 1e
Hamiltonian.
One surprising result shows that in general the interacting eigenenergies
are bounded between consecutive non--interacting 
two--electron (2e) energies, so that
the energy cost due to the interactions is usually much smaller than the
average repulsion $\langle U \rangle$, asigned to each pair in
the Coulomb blockade approach \cite{coulomb}.


Our study is also relevant for quantum
dots \cite{review}. A quantum dot coupled to electrodes has been modeled by one
impurity ($I=1$) on a one dimensional (1D), $N$--site chain
\cite{ng}. This represents a special case of
the Anderson model \cite{anderson},
with a momentum--dependent hybridization between the impurity and 
the conductance band. Unlike the case of a momentum--independent  hybridization,
we find that the behavior of the two electrons
on the ``dot" has a rich phase diagram, as function of
the ``dot" site energy $\epsilon_0$
and the hybridization, i.e., the hopping energy 
$t_0$ between the ``dot"
and the leads. These parameters can be tuned experimentally, by varying the
voltage on a gate coupled capacitively to the dot and the barriers between
the dot and the leads \cite{dotexp}.
We also find a delocalization transition at large $U$, for sufficiently 
negative $\epsilon_0$.
Our method allows a similar analytic treatment of the $I=1$ case
in higher dimensions.

More complex examples include a cluster of $I$ impurities coupled to
1D leads,
representing one large dot \cite{berkowitz}, 
two separate impurities, representing
double quantum dots (or artificial ``molecules") \cite{double}, etc.
The latter case may also shed light on the nature of the 
low--lying
states in the impurity band of doped semiconductors, which involve
``molecules" of impurity--pairs \cite{mott}. 
For $I=2$, the ``molecule" adjusts to our bound on the interaction energy
by crossing over to the Mott antiferromagnetic (singlet) state, 
in which the electrons are localized
on separate ions \cite{mott,harrison}.
Finally, the case $I=N$
represents the 2e
Hubbard model, whose non--random version ($\epsilon_i=0,\ U_i \equiv U$)
can be solved analytically in general dimensions using our method. 

We start with the 1e
Hamiltonian
\begin{equation}
{\cal H}_0= \sum_{\langle n,m \rangle}(t_{nm}|n \rangle\langle m|
+h.c.)+ \sum_{i=1}^I \epsilon_i |i \rangle \langle i|,
\label{H0}
\end{equation}
where $|i \rangle$ is a (spin--independent) state fully localized
on site $i$, the first sum runs over
all the site pairs in the system (including the impurities) and the second
sum runs over the $I$ impurities.
We first find the 1e eigenstates 
$|a \rangle \equiv \sum_n \phi_a(n)|n \rangle$ 
and eigenenergies $\epsilon_a$ of ${\cal H}_0$. 
This is usually
easy, being a linear problem (in any case, this involves at most the
diagonalization of an $N \times N$ matrix).
The on--site interactions take place only on the impurities,
\begin{equation}
{\cal H}_{\rm int}= \Sigma_{i=1}^I U_i\hat n_{i,\uparrow} \hat n_{i,\downarrow},
\label{Hint}
\end{equation}
where $\hat n_{i \sigma} \equiv |i \sigma \rangle \langle i \sigma |$ 
is the number operator of 
electrons with spin $\sigma=\uparrow,\downarrow$ on site $i$.
It is convenient to construct a basis for the 2e Hilbert 
space from the eigenstates of ${\cal H}_0$. These are split into
singlet ($S$) and triplet ($T$) spatial states,
both with energies $\epsilon_{ab} \equiv \epsilon_a+\epsilon_b$:
\begin{eqnarray}
|ab\rangle_S& = &(|a(1) \rangle |b(2) \rangle +|a(2) \rangle |b(1) \rangle)
\zeta_{ab},
\nonumber\\
|ab\rangle_T& = &(|a(1) \rangle |b(2) \rangle -|a(2) \rangle |b(1) \rangle)/
\sqrt{2},
\label{base}
\end{eqnarray}
where $|a(j) \rangle$ represents the 1e eigenstate 
of ${\cal H}_0$ for electron $j$,
$\zeta_{ab}=2^{-(1+\delta_{ab})/2}$ (with the
Kronecker $\delta$),
and the T states are used only for $a \ne b$.
To avoid double counting, 
we consider only states with $\epsilon_a \le \epsilon_b$.

Using this basis,
the matrix elements of ${\cal H}_{\rm int}$ involving T states
vanish for the on--site interaction of
Eq. (\ref{Hint}), and the energies of the T
states remain $\epsilon_{ab}$. We hence focus on the
S states, and omit the subscript S.
The nonzero S--S
matrix elements are 
$\langle ab|{\cal H}_{\rm int}|cd \rangle
= \Sigma_{i=1}^I U_i\eta_{ab}(i)^*\eta_{cd}(i)$,
where 
$\eta_{ab}(i)=2\zeta_{ab}\phi_a(i)\phi_b(i)$.
We now write the eigenfunctions of ${\cal H}_0+
{\cal H}_{\rm int}$, with energy $E$, as
$\Sigma_{\epsilon_a \le \epsilon_b} x_{ab}|ab\rangle$.
The coefficients must obey
\begin{equation}
(E-\epsilon_{ab})x_{ab}= \Sigma_i \sqrt{U_i}\eta_{ab}(i)^*A_i,
\label{xab}
\end{equation}
where $A_i \equiv \Sigma_{\epsilon_c \le \epsilon_d}
\sqrt{U_i}\eta_{cd}(i)x_{cd}$.
Using (\ref{xab}) we find
\begin{equation}
A_i= \Sigma_{j=1}^I S_{ij}A_j,
\label{AA}
\end{equation}
where
\begin{equation}
S_{ij}(E)=\sqrt{U_iU_j}\sum_{\epsilon_a \le \epsilon_b}\eta_{ab}(i)\eta_{ab}(j)^*
/(E-\epsilon_{ab}).
\label{SS}
\end{equation}
$S_{ij}/\sqrt{U_iU_j}=G(jj;ii;E)
\equiv \langle jj|(E-{\cal H}_0)^{-1}| ii\rangle$
is the non--interacting
2e Green's function in which the two electrons are on the same site,
cf. \cite{vonoppen}.

In addition to degenerate solutions \cite{com}, the new eigenstates are found by
requiring a non--zero solution to Eq. (\ref{AA}), namely that the 
$I \times I$ determinant $D(E) \equiv ||S_{ij}- \delta_{ij}||$ vanishes.
One way to find the new eigenvalues $\{E\}$ is to find
the $I$ eigenvalues ${\cal S}_i(E)$ of the matrix $S$,
and then solve the equations ${\cal S}_i(E)=1$.
The $A_i$'s are then given by the 
eigenvectors related to ${\cal S}_i(E)$,
and the $\{x_{ab}(E)\}$'s are found (up to a normalization constant)
from Eq. (\ref{xab}).
We emphasize that, unlike a perturbative expansion, this formalism
gives {\it exact} 
values for $E$ even for very large $U_i$'s. As we show below, this
enables us to find interesting transitions in that limit.

The equation $D(E)=0$ can be used to find bounds on the energies $\{E\}$.
Since each $S_{ij}$ has poles at every $\epsilon_{ab}$, we generally expect
each ${\cal S}_i$ also to have such poles. Although the 
residues of some of these
poles may vanish in special circumstances   
(e. g., ${\cal S}_i = S_{ii}=U_i/(E-2\epsilon_i)$ for isolated impurities,
$t_{nm} \equiv 0$, or for special symmetric cases, see below), this is not
expected to happen in the general random case, which we discuss now. 
When $E$ is very close to a non--interacting energy $\epsilon_{ab}$, one has
$S_{ij} \approx \sqrt{U_iU_j}\eta_{ab}(i)\eta_{ab}(j)^*/(E-\epsilon_{ab})$. 
It is then easy to show that in this approximation, $D(E) \approx
(-1)^I(1-{\rm Tr} S) \approx (-1)^{I} [1-\langle U \rangle_{ab}
/(E-\epsilon_{ab})]$,
with $\langle U \rangle_{ab} \equiv \sum_i U_i |\eta_{ab}(i)|^2$. 
Indeed, for very small $U_i$ the equation $D(E)=0$
reproduces the lowest
order perturbation result, $E \approx \epsilon_{ab}+\langle U \rangle_{ab}$.
As $E$ crosses through $\epsilon_{ab}$, $(-1)^{I+1}D(E)$ jumps from $-\infty$
to $\infty$. As $E$ increases between two consecutive non--interacting 
eigenvalues, $(-1)^{I+1}D(E)$
has no singularities, and thus varies smoothly from $\infty$ to $-\infty$.
Except for the special cases mentioned above, for which $D(E)$ splits
into products involving subspaces of the levels (see below), we thus 
conclude that $D(E)=0$ must have at least one (and up to $I$) solution(s)
between every pair
of such consecutive energies, and the new energies maintain the sequence
of the non--interacting ones. Specifically, if the 1e lowest levels are
$\epsilon_g$ and $\epsilon_u$ then the non--interacting lowest energies
are $\epsilon_{gg}=2\epsilon_g$ and $\epsilon_{gu}=\epsilon_g+\epsilon_u$, 
and the interacting ground
state energy $E_{gg}$ obeys $\epsilon_{gg}<E_{gg}<\epsilon_{gu}$.
Thus, the effective interaction cost for adding
the two electrons, $\Delta_{gg}=E_{gg}-\epsilon_{gg}$, is now bounded
by $(\epsilon_u-\epsilon_g)$, and is always smaller than 
$\langle U \rangle$.
For small $(\epsilon_u-\epsilon_g)$, this cost
is negligible.
Similar results apply
to all the levels except
the largest one (see below).
This renormalization of the interaction energy has direct consequences for the
issue of the Coulomb blockade, where one assumes that $\Delta_{ab} \approx 
\langle U \rangle_{ab}$, usually assumed to be independent of $ab$. 
It is interesting to note that small (and not evenly spaced)
values of $\Delta$ were observed in
a series of 2D quantum dots \cite{ashoori}.
It is tempting to relate these observations
to our result.

The simplest example concerns one impurity (or ``dot") on a 1D wire, closed with
periodic boundary conditions. We place the impurity at site $i=0$, with
energy $\epsilon_0$ and with real matrix elements $t_{0,1}=t_{0,N-1}
\equiv t_0$.
The other nearest neighbor matrix elements are set at $t_{n,n+1} \equiv t=1$,
for $n=1, 2, ...,N-2$, and all energies are scaled by $t$. 
The eigenenergies of ${\cal H}_0$ have the form $\epsilon_k=2 \cos(k)$.
Out of these, $N/2$ belong to odd states, of the form 
$\phi_k(n) \propto \sin(k n)$,
with $k=2 \pi\ell/N$, $\ell=1,2,...,N/2$ (for convenience, $N$ is even). 
These states do not ``feel" the impurity ($\phi_k(0)=0$),
and remain unchanged for all $\epsilon_0$ and $U$.
The remaining $N/2$ states are even functions of $n$, with
$\phi_k(n)= \phi_k(0)t_0 \cos[k(N-2n)/2]/\cos(k N/2)$ for $n \ne 0$,
where $|\phi_k(0)|$ is determined by $\sum_n|\phi_k(n)|^2=1$.
The allowed values of $k$ are given by
\begin{equation}
2 \gamma \tan(k N/2) \sin(k)=2(1-\gamma)\cos(k)-\epsilon_0,
\label{kk}
\end{equation}
where $\gamma \equiv (t_0/t)^2=t_0^2$.
An analysis of this equation yields the surprising phase diagram shown 
by the full lines in 
Fig. 1: In region A all the states are delocalized ($k$ is real). In region C
(or D+F) there exists one bound state above (or below) the conduction
band, with a localization length $1/\kappa_+$ and energy $\epsilon_+=
2 \cosh(\kappa_+)>2$, 
(or $1/\kappa_-$ and $\epsilon_-=-2 \cosh(\kappa_-)<-2$), where
$e^{\kappa_\pm}= \pm \epsilon_0/2 +\sqrt{(\epsilon_0/2)^2-1+2\gamma}$. 
Finally, both bound states exist in region B.
It is interesting to note that when
$\epsilon_0$ is inside the original conduction band, $|\epsilon_0|<2$,
the state on the impurity
becomes delocalized for any infinitesimal hybridization $\gamma$, 
and the impurity does not imply localization (as might be anticipated
in 1D). Furthermore, this state
becomes localized for {\it large} $\gamma$.
This surprising result arises since a larger $\gamma$ implies a larger 
repulsion of the localized state below the band.

Combining the normalization condition with Eq. (\ref{kk}) yields
(for $N \gg 1$)
$|\phi_b(0)|^2=1/[1+2\gamma/(e^{2\kappa}-1)]$ for the bound states $b=\pm$, and
$|\phi_k(0)|^2=2\pi f(k)/N$, with
$\pi f(k)=4\gamma\sin^2(k)/[4\gamma^2\sin^2(k)+(2(1-\gamma)\cos(k)
-\epsilon_0)^2]$. The latter result implies that unless one has a uniform
chain, where $\gamma=1$ and $\epsilon_0=0$, the weight of the states with
$k$ at the band edges $0$
and $\pi$ vanishes on the impurity. This will turn out to
be crucial below.

For $I=1$, Eq. (\ref{AA}) reduces to a single equation,
$D(E)=Us(E)-1$, and the eigenenergies $E$ obey
$s(E)=1/U$.   
Similar equations are encountered in numerous 
cases, e. g., in the Cooper 2e problem or in Kohn's model
of the insulating state\cite{eco}.
Indeed, $s(E)$ jumps from $-\infty$ to $\infty$ as $E$ crosses
each non--interacting energy $\epsilon_{ab}$,
and we find one new eigenvalue between every pair of non--interacting
energies, as described above. In regions B, D and F, $\epsilon_g=\epsilon_-$
and $\epsilon_u=-2$ (the bottom of the band). Thus,
$\Delta_{gg}$ is bound by $|\epsilon_-|-2$, and is always smaller than $U$.
In fact, $\Delta_{gg} \rightarrow 0$ when the 1e bound state approaches
the band, at the line $\epsilon_0=2(\gamma-1)$.
For another insight into the smallness of $\Delta_{gg}$,
rewrite $s=|\phi_-(0)|^4/(E-2\epsilon_-)-\Gamma(E)$. For $E<\epsilon_{gu}=
\epsilon_--2$, one has $\Gamma>0$, and
$|\phi_-(0)|^4/(E-2\epsilon_-)=1/U_{\rm eff}$, with the smaller
{\it renormalized}
repulsive energy $U_{\rm eff}=U/(1+\Gamma U)<U$.
It is interesting to note that in regions D and F, 
$\Gamma$ is proportional to $\gamma$,
and therefore the renormalization of $U$ increases with the hybridization.
The Coulomb blockade picture is restored for $\gamma \ll 1$.

The details of $s(E)$ depend on the parameters $\epsilon_0$
and $\gamma$. 
To treat the bands (of states $|kk' \rangle$ and $|\pm k \rangle$) in the limit
$N \rightarrow \infty$, we replace the summations by integrals:
\begin{eqnarray}
s&=&\sum_{b,b'} \frac{2^{1-\delta_{bb'}}
|\phi_b(0)\phi_{b'}(0)|^2}{E-\epsilon_b-\epsilon_b'}
\nonumber\\
&+&2\sum_b |\phi_b(0)|^2 \int_0^\pi dk \frac{f(k)}{E-2\cos(k)-\epsilon_b}
\nonumber\\
&+&2 \int_0^\pi dk\int_0^\pi dk'\frac{f(k)f(k')}
{E-2\cos(k)-2\cos(k')},
\label{S00}
\end{eqnarray}
where the bound states $b,b'=\pm$ are included only when they exist
(regions C, D, F and B).
For $E$ inside the band we must return to the discrete sum, and the dense
energies practically don't shift. 

Our most interesting results arise in region D+F.
Here we have only one bound state, $b=-$, so that $s(E)$
is negative for $E<\epsilon_{gg}=2\epsilon_-$, 
and decreases from $+ \infty$ as $E$
increases above this value, towards the band which starts at
$\epsilon_{gu}=\epsilon_--2$. For finite $N$, $s$ would diverge towards $-\infty$
as $\epsilon_{gu}$ is approached, implying a persistent bound solution with a
discrete energy below the band for all $U$. This would also be the case
for the continuum case, $N \rightarrow \infty$, if one had a non--zero
value of $f(0)$ (from $|\phi_k(0)|^2$ at $k=0$), due to the divergence of the
1D density of states there. However, as noted above, $f(k)$ vanishes at $k=0$,
yielding a finite value $s(\epsilon_{gu}) \equiv s_c$.
We studied $s_c$ as function of $\epsilon_0$ and $\gamma$, and
found that in region D of Fig. 1 one has $s_c<0$, so that the equation
$s(E)=1/U$ still has a discrete bound (``insulating") state there.
However, in region F one has
$s_c>0$, 
implying a disappearance
of this bound state for
$U>U_c=1/s_c$. This transition is intuitively
easy to understand: when $\gamma$ is very small, this transition occurs
when the energy of the two electrons in the isolated atom,
$2\epsilon_0+U$, exceeds that of the state in which one electron remains
bounded while the other moves to the band, which is equal to $\epsilon_0-2$.
At finite $\gamma$, a larger negative $\epsilon_0$ implies a 
smaller localization length $1/\kappa_-$. The electrons are then more localized on the impurity,
and $\Delta_{gg}$ is larger, leading to the
``insulator to metal" transition from region D to region F.
It should be noted that although the bound state is a singlet, 
with total spin zero, the new ground
state in region F has one bound electron and one ``free" electron. Such a
state does not feel the e--e repulsion, and is thus practically 
degenerate with the slightly lower triplet state (for large $N$, the difference
is of order $1/N$). 
Unlike the ``insulator" singlet (or ``antiferromagnetic")
ground state, which has no net magnetic moment,
this ``metallic" state in region F is {\it paramagnetic}. 
This difference should
be measurable in an external magnetic field.

It is interesting to study the crossover from the ``mesoscopic" case, of finite
$N$, to the thermodynamic limit discusssed above. 
Firstly, for finite large $N$ the transition from region A to region D
is smeared, occuring when $(\epsilon_{-}+2)$ becomes comparable
to the spacing between the band states, of order $1/N^2$.
Secondly, in the mesoscopic case,
$s(E)$ diverges to $-\infty$ as $E$ approaches the lowest band state
(slightly above $\epsilon_{gu}$). However, in region F, $s(E)$
first gets very close
to $s_c$, and only then drops sharply 
to $-\infty$. For $U>U_c$, the resulting
``bound" state will thus occur very close to $\epsilon_{gu}$, making it almost
indistinguishable from the states inside the band.
This implies a ``smeared" transition from region D to region F for large
mesoscopic systems.

Similar interesting effects occur above the band, in regions A, D and F.
The highest band energy is $E=4$, corresponding to the
upper limit of the last term in Eq. (\ref{S00}). Since $f(\pi)=0$, $s(E)$
remains finite as $E$ approaches $4$ from above,
and $0< s(E=4) \equiv s_x < \infty$
[$s_x=\infty$  only on the boundary $\epsilon_0=2(1-\gamma)$]. 
As $E$ increases from $4$ to
$\infty$,
$s$ decreases from $s_x$ to zero, implying a discrete solution to
the equation $s(E)=1/U$,
with $E>4$ above the band, if
$U>U_x=1/s_x$. We found (by calculating 
$x_{ab}$) that the corresponding
wave function decays  away from the impurity.
Since this new state is {\it above} the band, it describes an
``antibound"  2e excitation, which is reminiscent of the
upper Hubbard band at low concentration \cite{pwa}.
Our analysis finds a phase transition
of this excited state, from an insulating localized
state at $U>U_x$ to a conducting state for smaller $U$.
This transition also becomes ``smeared" for finite $N$.

Now we give a very brief summary of our results for $I=2$, on the 1D ring.
Consider the very symmetric case, with
 the two impurities at sites $i=i_1,\ i_2$
having $\epsilon_i \equiv \epsilon_0$, $U_i \equiv U$, $t_{i,i \pm 1}=t_0$
and $i_2-i_1=R$.
The 1e real wave functions separate into even and odd
ones, with $\phi_a(i_1)=\pm \phi(i_2)$.
Depending on $\epsilon_0$, $t_0$ and $R$ one again finds a rich 
1e phase diagram. In particular, the low energy
1e states may start with the band, or have a bound even
(``bonding")
state or (above a minimal value of $R$, $R_x$) have also an odd bound 
(``antibonding") state between the
bonding state and the band.
The high symmetry implies
that $S_{11}=S_{22}$ and $S_{12}=S_{21}$,
with eigenvalues ${\cal S}_\pm (E)=S_{11} \pm S_{12}$.  
Writing ${\cal S}_\pm (E)$ as sums over non--interacting singlet states,
we find that ${\cal S}_+$ contains only the even--even and the odd--odd states,
while ${\cal S}_-$ contains only the even--odd states.
Thus, $D(E)=({\cal S}_+-1)({\cal S}_--1)$,
and the problem splits into two separate spaces.
Consequently, $\Delta_{gg}$ is now bounded by the first non--interacting
excited state {\it within the 
even--even and odd--odd subspace}.
The analysis of each equation ${\cal S}_\pm =1$ is now similar
to that of $I=1$, and the resulting phase diagram again contains delocalization
and magnetic transitions \cite{details}.
It is possible to follow the coefficients $\{x_{ab}\}$, and see 
exactly how $U$ 
causes a gradual crossover from the ``molecular" to the ``atomic" 
limits as $R$ increases \cite{harrison}. In the latter localized states,
the effective interaction energy obeys our bound by being of order $1/U$. 
This crossover would be missed for $R<R_x$, if one
were to ignore the hybridization with the band
(as commonly assumed
 \cite{tokura}).

Our method is easily generalized to higher dimensions and to larger $I$.
It should be particularly helpful in numerical work on random systems.
Another simple example concerns the non--random
Hubbard model, where $I=N$. The matrix
$S_{ij}$ is easily diagonalized in Fourier space, yielding the upper 
Hubbard band. This and other examples will be published elsewhere
 \onlinecite{details}.

We acknowledge many discussions with Yehoshua Levinson, and comments from
Peter W\"{o}lfle. This research is supported by grants from
the Israel Science Foundation
and the French Ministry of Research and Technology.

\vspace{2cm}
\begin{figure}
\centerline{\psfig{figure=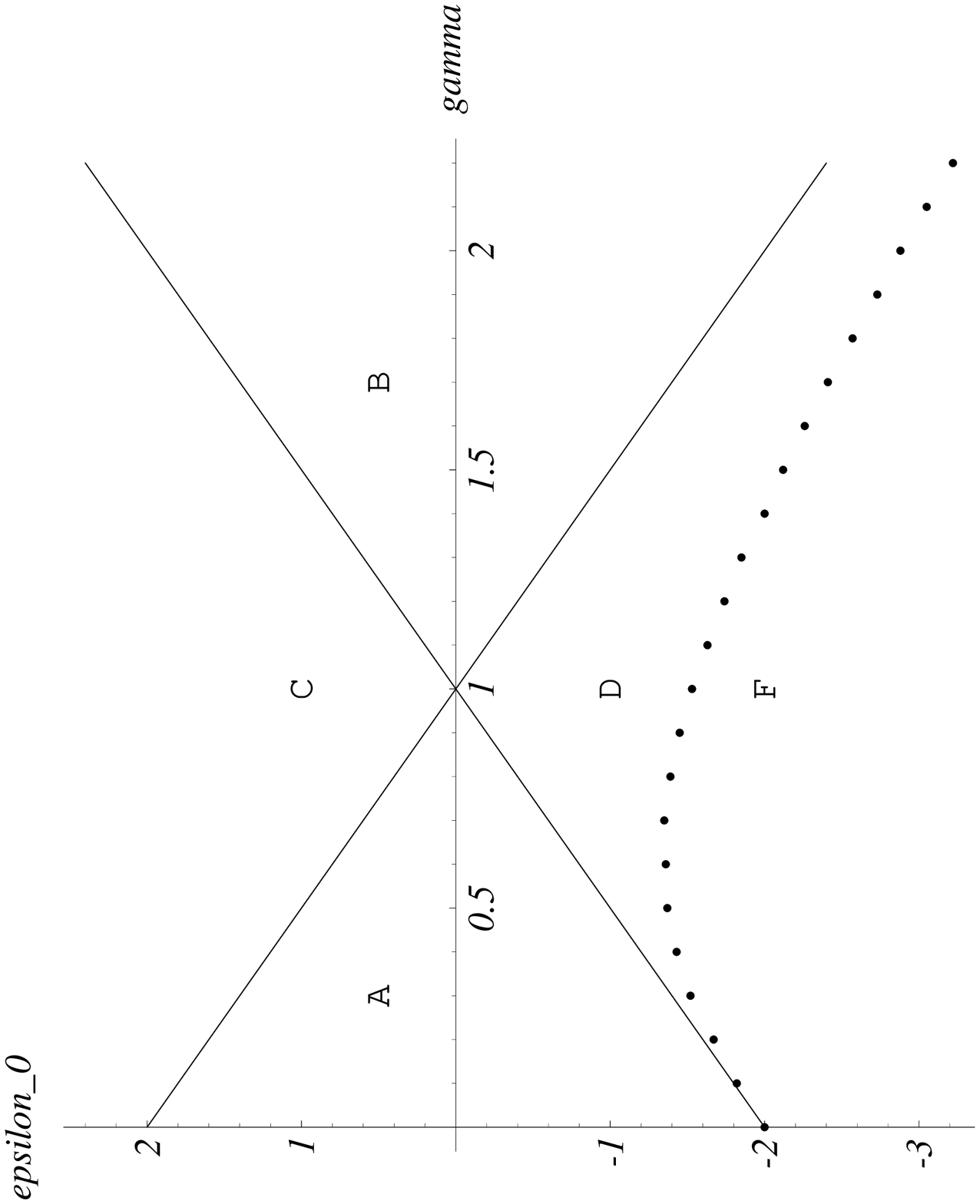,width=8cm}}
\caption{$\epsilon_0-\gamma$ phase diagram for the single impurity case.}
\label{fig1}
\end{figure}


\begin{references}
\bibitem{com}If the energy $\epsilon_{ab}$ is $M-$fold degenerate 
then one can have
up tp $(M-1)$ solutions for which all $A_i=0$, $E=\epsilon_{ab}$
remains unaffected by the interactions and $x_{a,\alpha;b,\beta}$
are the solutions of $\sum_{\alpha,\beta} \eta_{a,\alpha ; b,\beta}(i)
x_{a,\alpha ; b,\beta}(i)=0$, where $\alpha, \beta=1,...,M$ denote the
degenerate states.
\end{references}
\end{document}